\documentclass[preprint,floats,aps,epsfig,nofootinbib,amssymb]{revtex4-1}
\usepackage{mathrsfs}

\usepackage{slashed}
\usepackage{graphicx,color}
\usepackage{dcolumn}
\usepackage{bm}
\usepackage{subfig}
\usepackage{graphicx}
\usepackage{amssymb}

\begin{document}
\preprint{ACFI-T14-26}

\title{First order electroweak phase transition triggered by the Higgs portal vector dark matter}

\author{Wei Chao}
\email{chao@physics.umass.edu}

\affiliation{Amherst Center for Fundamental Interactions, Department of Physics, University of Massachusetts-Amherst
Amherst, MA 01003 
 }

\vspace{3cm}

\begin{abstract}

We investigate an extension of the Standard Model (SM) with a $U(1)^\prime$ gauge symmetry, which is spontaneously broken by a complex scalar singlet and  where the new gauge boson is a  stable dark matter candidate via a $Z_2$ flavor symmetry. The possibility of generating a strongly first order electroweak phase transition (EWPT) needed for the electroweak baryogenesis mechanism in this model is studied using a gauge independent method.  Our result shows a considerable parameter space where both successful dark matter phenomenologies and a strongly first order EWPT can be achieved.

\end{abstract}

\maketitle
\section{Introduction}
With the discovery of the Higgs-like scalar at the CERN LHC~\cite{atlas,Aad:2013wqa,cms,Chatrchyan:2013lba}, the Higgs mechanism~\cite{higgs}  for spontaneous breaking of the gauge symmetry in the standard model (SM) appears to be a correct description of nature. It opens a new era of direct probes of electroweak symmetry breaking.
It was observed by Kirzhnits and Linde~\cite{Kirzhnits:1972ut} that spontaneously broken symmetries are usually restored at the high temperature. Thus the broken electroweak symmetry is expected to be restored in the early Universe. 
A transition occurred about $10^{-10}$ second after the Big Bang.  The dynamics of the electroweak phase transition (EWPT), which are still open questions, are important in the attempts to explain the observed matter-antimatter asymmetry of the Universe in terms of baryon number violation in the electroweak theory and cosmological models of baryogenesis~\cite{Rubakov:1996vz,Morrissey:2012db}. In particular, the condition that the baryon excess generated at the EWPT will not be washed out requires a strong enough first order phase transition, which translates into an upper bound on the Higgs boson mass. The $125~{\rm GeV}$ Higgs boson will be too heavy to give rise to a first order EWPT.  Theoretically, dynamics of the EWPT are determined by the effective Higgs potential at the finite temperature, which are tightly connected with the Higgs interactions at the zero temperature. A strongly fist-order EWPT requires new Higgs interactions with particles  beyond the SM.   


EWPT is one of the necessary conditions for a workable electroweak baryogenesis mechanism of generating the matter-antimatter asymmetry, which results in the visible part of our Universe. For the invisible part of our Universe, precisely cosmological observations have confirmed the existence of the non-baryonic cold dark matter $\Omega h^2 =0.1186\pm0.0031 $~\cite{Ade:2013zuv}, which provides another evidence of the new physics beyond the SM. Much effort has been employed to interpret the dark matter signals. Among various possible dark matter candidates that have been explored in the literature, the weakly interacting massive particle(WIMP)~\cite{dmn1,dmn2,dmn3}  stands out as the most interesting scenario. However the nature of the dark matter and  the way that the dark matter interacts with the ordinary matter are still  mysteries.  The discovery of the Higgs boson opens up new ways of probing the dark matter.  It is natural to consider the Higgs portal dark matter model~\cite{Patt:2006fw,Kim:2006af,MarchRussell:2008yu,Kim:2008pp,Ahlers:2008qc,Feng:2008mu,Andreas:2008xy,Barger:2008jx,Kadastik:2009ca,Piazza:2010ye,Arina:2010an,Kanemura:2010sh,Englert:2011yb,Low:2011kp,Djouadi:2011aa,Kamenik:2012hn,Gonderinger:2012rd,Lebedev:2012zw,LopezHonorez:2012kv,Okada:2012cc,Djouadi:2012zc,Bai:2012nv,Englert:2013gz,Bian:2013wna,Chang:2013lfa,Khoze:2013uia,Okada:2013bna,Fedderke:2014wda,Baek:2012se,Lebedev:2011iq,Chao:2012pt} , in which dark matter couples to the SM Higgs in the form ${\cal O}_{\rm DM} {\cal O}_{\rm Higgs}$, where ${\cal O}_{\rm Higgs}$ is the Higgs bilinear, $H^\dagger H$, which is one of the lowest mass dimension and gauge invariant operators in the SM. ${\cal O}_{\rm DM}$ is the dark matter bilinear and can be written as $\phi_{\rm DM}^\dagger \phi_{\rm DM}$  for the scalar dark matter, $1/\Lambda \bar\chi_{\rm DM} \chi_{\rm DM}~{\rm and (or) }~1/\Lambda\bar\chi_{\rm DM} i\gamma^5 \chi_{\rm DM}$ for the fermonic dark matter, where $\Lambda$ is roughly the mass scale of the mediators for a ${\cal O} (1)$ coupling between the dark matter and  the SM Higgs, $V_\mu^\dagger V_\mu^{}$ for the vector dark matter.

In this paper we study the possibility of getting a strongly first order EWPT in the framework of the Higgs portal vector dark matter model~\cite{Kanemura:2010sh,Lebedev:2011iq,Djouadi:2011aa,Baek:2012se}, which extends the SM with a spontaneously broken $U(1)^\prime$ gauge symmetry and a $Z_2$ discrete flavor symmetry, that stabilizes the new vector field as a dark matter candidate.  A complex scalar singlet is needed to break  the $U(1)^\prime$ gauge symmetry and gives the mass to the vector dark matter.  The scalar singlet is mixed with the SM Higgs via the quartic interaction.  The strongly first order EWPT is triggered by the same interaction in our model. For the effect of scalar singlets on the EWPT, see~\cite{Pietroni:1992in,Espinosa:1993bs,John:1998ip,Huber:2000mg,Ham:2004cf,Cline:2012hg,Cline:2013gha,Katz:2014bha,Alanne:2014bra,Profumo:2007wc,Noble:2007kk,Espinosa:2011ax,Profumo:2014opa,Curtin:2014jma}. Our study is new in the following two aspects: 
\begin{itemize}
\item We treat the effective potential in a gauge invariant way. The critical temperature $T_C$ and the energy scale $\bar v(T)$ is gauge invariant; \item EWPT is closely related to the phenomenology of the vector dark matter in our model.
\end{itemize}
Our study shows that even though there are strong constraints on the model from the exclusion limits of the dark matter direct detection  experiments  such as LUX~\cite{Akerib:2013tjd}, one can still find the parameter space, where all the dark matter constraints can be satisfied and  a strongly first order EWPT can be generated. 

The paper is organized as follows: In section II we give a brief introduction to the model. Section III is the study of the dark matter phenomenology. We investigate the EWPT and its correlation with the Higgs portal vector dark matter  in section IV. The last part is the concluding remarks.  

\section{The model}

We assume the dark matter is a vector boson, $V_\mu$,  which can be the gauge field of a $ U(1)^\prime$ gauge symmetry that is spontaneously broken.  SM fields carry no $U(1)^\prime$ charge. The only field charged under the $U(1)^\prime$ is a complex scalar singlet whose vacuum expectation value (VEV) breaks the new gauge symmetry spontaneously and gives rise to a non-zero mass of $V_\mu$.  The model has a $ Z_2$ discrete flavor symmetry, under which $V_\mu$ is odd and all the other fields are even, which makes $V_\mu$ a stable dark matter candidate. The relevant Lagrangian can be written as
\begin{eqnarray}
{\cal L} = (D_\mu H)^\dagger (D_\mu H) +(D_\mu^\prime S)^\dagger (D_\mu^\prime S) -V(H,S) \label{lagrangian}
\end{eqnarray}
where 
\begin{eqnarray}
D_\mu^\prime =\partial_\mu - ig_N Q_S  V_\mu
\end{eqnarray}
with $g_N$ being the gauge coupling of the $U(1)^\prime$ and $Q_S$ the $U(1)^\prime$ hypercharge of the scalar singlet. We take $Q_S=1 $.  The Higgs potential is  
\begin{eqnarray}
V(H, S) =  - \mu^2 H^\dagger H - \mu_s^2 S^\dagger S + \lambda  (H^\dagger H)^2 + \lambda_1 (S^\dagger S)^2  +\lambda_2 (H^\dagger H) (S^\dagger S) \label{potential}
\end{eqnarray}
where $H= (h^+, ~(h+i A + v) /\sqrt{2})^T$and $S = (s + i \rho + v_s^{}) /\sqrt{2}$. After imposing the conditions of the global minimum, one has
\begin{eqnarray}
v^2\approx { 2 \mu_s^2 \lambda_2^{} - 4 \mu^2 \lambda_1 \over \lambda_2^2 -4 \lambda_1^{} \lambda }  \; ,\hspace{1cm}
v_s^2 \approx { 2\mu^2 \lambda_2 - 4  \mu_s^2 \lambda \over \lambda_2^2 - 4 \lambda \lambda_1^{} } \; , \label{vev}
\end{eqnarray}
$A$ and $\rho$ are goldstone bosons eaten by $W^3_\mu$ and $V_\mu$ respectively.  The mass matrix of the CP-even scalars is
\begin{eqnarray}
M_{\rm even}^2 =\left(  \matrix{ 2 \lambda v^2 & \lambda_2 v v_s \cr \lambda_2 v v_s & 2 \lambda_1 v_s^2 }\right)
\end{eqnarray}  
which can be diagonalized by a $2\times 2 $ orthogonal unitary matrix.  Mass eigenvalues  and the mixing angle can be written as
\begin{eqnarray}
m_{1,2}^2 &=& (\lambda v^2 + \lambda_1 v_s^2 ) \pm \sqrt{ (\lambda v^2 -\lambda_1 v_s^2 )^2 + (\lambda_2 vv_s)^2 }  \; , \label{eigenvalue}\\
\theta &=& {1\over 2}  \arctan  \left( \lambda_2 v v_s \over \lambda v^2 -\lambda_1 v_s^2  \right)\; .
\end{eqnarray}
Assuming  $m_s$ is heavier, one has $m_s^2 =m_h^2 +2 \sqrt{ (\lambda v^2 -\lambda_1 v_s^2 )^2 + (\lambda_2 vv_s)^2}$, where $m_h$ is the mass of the SM Higgs.

The scalar singlet decays into the new vector boson as well as SM fields via the mixing with the SM Higgs, the decay rate can be written as
\begin{eqnarray}
\Gamma_s &=&\sum_{\alpha=V,W,Z} { f_\alpha^2 m_\alpha^4  \sqrt{m_s^2 -4 m_\alpha^2 }\over  4 (1+\delta_\alpha) \pi v_s^2 m_s^2 } \left(  3 -{m_s^2 \over m_\alpha^2 } + {m_s^4 \over 4 m_\alpha^4}\right) \theta (m_s -2 m_\alpha)    \nonumber \\&
+ &\sum_{\psi=b, t} {  s_\theta^2 m_\psi^2 (m_s^2 - 4 m_\psi^2)^{3/2}   \over 8 \pi v^2 m_s^2 } \theta (m_s- 2 m_\psi)
\end{eqnarray}
where $s_\theta =\sin \theta $, with $\theta $ the mixing angle between the SM Higgs and the scalar singlet, $f_V =\cos \theta $ and  $ f_{W, Z} = (v_s/v)  \sin \theta $, $\delta_{V,Z}=1$ and $\delta_W=0$. If $m_s < m_h$, $s$ mainly decayes into $b \bar b $ and $\tau \bar \tau$.

\section{Dark Matter Phenomenology}
In order to determine the relic density of the dark matter and the source function of cosmic-ray particles derived from the dark matter annihilation in the Galactic halo, which is relevant to the dark matter indirect detection, one needs to the calculate the the dark matter annihilation.
The results are given by
\begin{eqnarray}
\sigma(V_\mu V_\mu\to \bar f f ) & = & {c_\theta^2 s_\theta^2 \over 18 \pi  s } \left( { m_V^2 m_f^{} \over v_s v_h } \right)^2  {  (s- 4 m_f )^{3/2 } \over  \sqrt{s -4 m_V^2 } }  {\cal T } \times {\cal P } \; , \\
\sigma(V_\mu V_\mu \to GG ) & = &  {c_\theta^2 s_\theta^2 \over 9 (1+\delta_G) \pi  s }  \left( { m_V^2 m_f^{} \over v_s v_h } \right)^2  {\sqrt{s -4m_G^2  \over s- 4 m_V^2 } } \left( 3- {s \over m_G^2 } + { s^2 \over 4 m_G^4 }\right)  {\cal T } \times {\cal P }  \; , \\
\sigma(V_\mu V_\mu \to ss) &\approx&{ c_\theta^4 m_V^4  [ 6 c_\theta^2 \lambda_1 v_s^2 + (s-m_s^2 )]^2\over 72 \pi s v_s^4  [(s-m_s^2 )^2 + m_s^2 \Gamma_s^2 ]}  {\sqrt{s- 4m_s^2 \over s-4 m_V^2 }}  *{\cal T }  \; , \\
\sigma(V_\mu V_\mu \to sh ) &\approx& {m_V^4 [ - 6 c_\theta^3  s_\theta \lambda_1 v_s^2 +c_\theta^4 \lambda_2 v_s v_h + c_\theta s_\theta (s-m_s^2 )]^2 \over 36 \pi s^{3/2 } v_s^4  [(s-m_s^2 )^2 + m_s^2 \Gamma_s^2 ]}  \sqrt{ \lambda(s, m_s^2 , m_h^2 ) \over s - 4 m_V^2 } {\cal T } \; .
\end{eqnarray}
where 
\begin{eqnarray}
{\cal  T  } =    3 - { s \over m_V^2 } + {s^2 \over 4 m_V^2 } \; , \hspace{1cm } {\cal P } =  { ( { m_h^2 - m_s^2 })^2 + (m_h \Gamma_h -m_s \Gamma_s)^2 \over [(s-m_s)^2 + m_s^2 \Gamma_s^2 ][(s-m_h^2)^2 + m_h^2 \Gamma_h^2 ] } \; .
\end{eqnarray}
where $s$ is the Lorentz invariant Mandelstam variable, $m_{h,s}$ and $\Gamma_{h,s}$ are masses and decay widths of $h$ and $s$ respectively, $m_f$ and $m_G$ are the masses of fermions and vector bosons in the final states.

A dark matter is detectable through its scattering on atomic nuclei on the earth, by production at particle colliders or through detection of its annihilation radiation in our galaxy.  Here we focus on the dark matter direct detection in the deep underground laboratories, which registers the interaction  of  through-going dark matter. The dark matter-quark effective Hamiltonian in our model can be written as
\begin{eqnarray}
H_{eff} = c_\theta s_\theta {2 m_V^2 \over v_s } V_\mu ^{}V^\mu \left({1 \over m_s^2 } -{1\over m_h^2 }\right) {m_q \over v } \bar q q \; .
\end{eqnarray}
Parameterizing the nucleonic matrix element as $\langle N |\sum_q m_q \bar q q |N\rangle =f_N m_N$, where $m_N$ is the proton or neutron masses and $f_N$ is the nucleon form factor, the cross section for the dark matter scattering elastically from a nucleus is given by
\begin{eqnarray}
\sigma= {c_\theta^2   s_\theta^2 \mu^2 \over \pi  } \left[ {m_V m_p f_p \over v v_s }  \left( {1 \over m_h^2 } -{1 \over m_s^2 }\right)\right]^2 
\end{eqnarray}
where $\mu = m_V m_p/(m_V+ m_p)$, which is the reduced mass of the dark matter and the proton, with $m_p$ being the proton mass.  We refer to \cite{dmn1,dmn2,dmn3} for explicit values of $f_{p}$.

\begin{figure}[h!]
\centering
\includegraphics[width=0.45\textwidth]{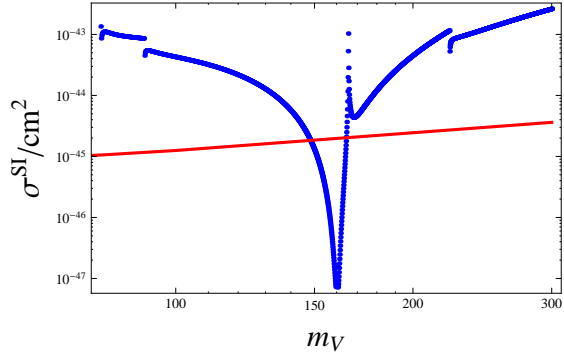}
\caption{
Spin independent dark matter nucleon scattering cross section as the function of the dark matter mass.  We set $m_s =320~{\rm GeV}$ and $v_s=1~{\rm TeV}$. All  the points in blue color give the correct dark matter relic density.  The red solid line is the  limit of LUX. 
 }\label{directt}
\end{figure}

For $m_V  < m_W$, the dark matter pair annihilate mostly into quark and lepton pairs, the amplitude of which is suppressed by the Yukawa couplings. As a result the dark matter relic abundance for $m_V < m_W$ will be too large to be consistent with the dark matter observations. For $m_W < m_V < m_h$, the dominate channels are $VV\to W^+ W^-$ and $VV \to ZZ$. When $m_V$ gets even bigger, $VV\to  hh, hs, ss$ are no longer kinematically forbidden and become dominant annihilation channels.
We plot in Fig. \ref{directt} the spin independent dark matter nucleon scattering cross section as the function of the dark matter mass by setting $m_s=320~{\rm GeV}$ and $v_s = 1~{\rm TeV}$. All the points in the blue curve give a correct dark matter relic density. The red solid line is the exclude limit given by the LUX~\cite{Akerib:2013tjd}.  One can conclude that the Higgs portal dark matter model  may survive only at the nearby of the resonance of the scalar singlet.   

\begin{figure}[h!]
\centering
\includegraphics[width=0.45\textwidth]{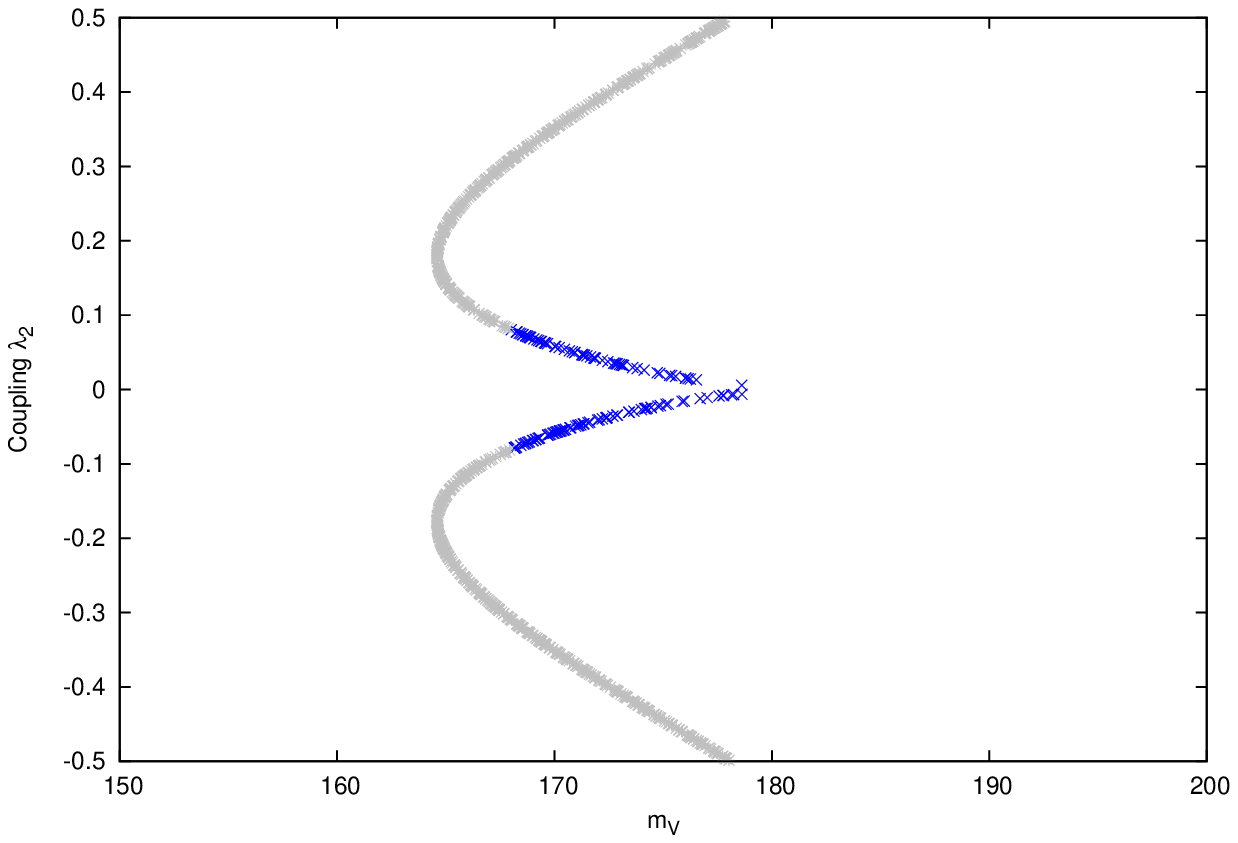}
\hspace{0.5cm}
\includegraphics[width=0.45\textwidth]{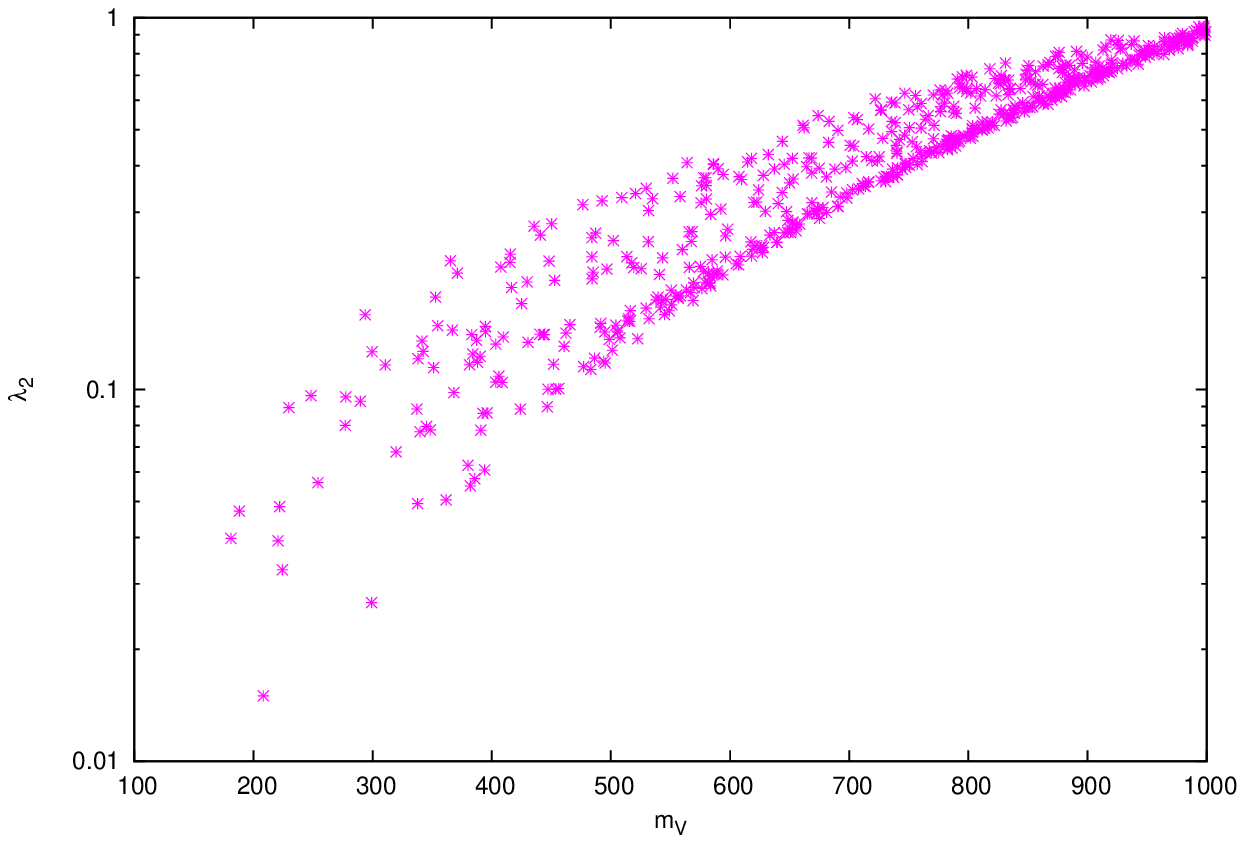}
\caption{
 $\lambda_2$ as the function of the dark matter mass. All the points give the correct dark matter relic density. Points in gray color are excluded by the LUX (left panel), points in blue (left panel) and magenta color (right panel) are allowed by the LUX. 
 }\label{directt2}
\end{figure}

We plot in the left panel Fig. \ref{directt2} the coupling $\lambda_2$ as the function of the dark matter mass by setting $\lambda_1=0.1$ and $v_s=1~{\rm TeV}$. All the points in the figure give a correct dark matter relic density.  Points in blue color satisfy the constraint of the LUX, while points in grey are excluded by the LUX. One can see that the dark matter direct detection puts  strong constraint on the coupling $\lambda_2$. It is worth to mention that the constraint changes as the initial inputs vary.  In the right panel of Fig. \ref{directt2}, we show the scattering plots in the $\lambda_2 -m_V$ plane, where each point satisfies all the dark matter constraints. We have set $v_s$ and $\lambda_1$ as free parameters varying in  ranges: $v_s\in[100,2000]~{\rm GeV}$ and $\lambda_1 \in(0,~2)$, when making the plot.

\section{Electroweak phase transition}
In this section we study the electroweak phase transition in the Higgs portal vector dark matter model. The Lagrangian was given in Eq. (\ref{lagrangian}) and (\ref{potential}). Fields contributing to the effective potential are the Higgs field, Goldstone bosons, gauge bosons, new scalar singlet, vector dark matter and the top quark. Field dependent mass squares are given in Table. \ref{massbg}. Thermal masses of the SM Higgs, the scalar singlet and the vector dark matter are given by
\begin{eqnarray}
\Pi _h &=&\left(  { { 3 g^2 + g^{\prime2 } \over 16} + {\lambda \over 2 } + {\lambda_2 \over 12 } + {h_t^2 \over 4} } \right)T^2 \; ,  \label{themalma}\\
\Pi_s &=&\left(  {g^{\prime \prime 2 } \over 4 }+ {\lambda_1 \over  3 } + {  \lambda_2 \over 6} \right) T^2 \; ,  \label{themalmb}\\
\Pi_{V}^L &=& {2 \over 3 }g^2 T^2   \; . \label{themalmc}
\end{eqnarray}
The effective potential, which is critical for the EWPT, can be written as
\begin{eqnarray}
V_{\rm eff} = V_0 +V_{\rm CW}+ V_T \; ,
\end{eqnarray}
where $V_{\rm CW}^{}$, known as Coleman-Weinberg potential, contains the one-loop contributions to the zero temperature effective potential,  $V_T$ includes the finite temperature contributions. Both $V^{}_{\rm CW}$ and $V_T$ receive contributions from particles that couple to the Higgs.  A particle's contribution to the effective potential is determined by its multiplicity, its fermion number and its mass in the presence of a background Higgs field.

The Coleman-Weinberg effective potential can be expressed in terms of the field dependent masses 
\begin{eqnarray}
V_{\rm CW} &=&  {1 \over 64 \pi^2 } \sum_i (-1)^{2 s_i }n_i m_i^4(h,s,\xi) \left[ \log { m_i^2 (h, s, \xi ) \over \mu^2 } -C_i  \right] \; ,  \label{b}
\end{eqnarray}
where $\mu $ is the renormalization scale, fixed to be $v_0$, the sum is over all fields that interact with the scalar fields, $n_i$ and $s_i$ are the number of degrees of freedom and the spin of the $i$-th particle. $C_i$ equals to $5/6$ for gauge bosons and $3/2$ for scalars and fermions. We calculate the effective potential in $R_\xi$ gauge.

\begin{table}[t]
\centering
\begin{tabular}{c|c||c|c||c|c}
\hline
\hline scalars &  masses  &  gauge fields & masses & fermions & masses \\
\hline
$\phi  $ & $ -\mu_h^2 +3 \lambda h^2 + {1\over 2 } \lambda_2 s^2 $ & $W$  &${ g^2 \over 4 } h^2 $ & $t $ & ${ h_t^2 \over 2 } h^2 $ \\
$\chi $ &  $   -\mu_h^2 + \lambda h^2 + {1\over 2 } \lambda_2 s^2 $ & $Z$  & ${g^2 + {g^\prime}^2 \over 4 } h^2 $&&\\
$\varphi$ & $ -\mu_s^2 + 3 \lambda_1 s^2 + {1 \over 2 } \lambda_2 h^2 $ & $\gamma$ & 0 &&\\
$\rho $ &  $ -\mu_s^2 +  \lambda_1 s^2 + {1 \over 2 } \lambda_2 h^2 $  & $V$ & $g^{\prime \prime 2 } s^2 $&&\\
\hline \hline
\end{tabular}
\caption{ Field-dependent masses of various particles.  }\label{massbg}
\end{table}

The temperature dependent effective potential can be calculated using standard techniques. It receives two contributions: the one-loop contribution and the bosonic ring contribution, which depends on thermal masses. Imposing renormalization conditions preserving the tree level values of VEVs and working in the $R_{\xi}$ gauge, the fields-dependent part can be written as
\begin{eqnarray}
V_T^{}&=&  {T^4 \over 2\pi^2 } \left\{\sum_{i\in B} n_i J_B\left [ {m_i^2(h,s,\xi) \over T^2 }\right] -\sum_{j\in F} n_jJ_F\left [ {m_j^2(h) \over T^2 }\right ] -\sum_{k\in G } n_k J_B\left [  {m_k^2(h,s,\xi) \over T^2 }\right ]\right\}\; , \label{c}
\end{eqnarray}
where the first term is contributions of bosons, the second term is contributions of fermions and the third term is contributions of ghosts.  The explicit expression of functions $J_{B(F)} (x)$ can be found in  Ref.~\cite{Quiros:1999jp}.  The ring contribution can be written as
\begin{eqnarray}
V_{T}^{\rm ring} = {T \over 12 \pi} \sum_i n_i \left\{  (m_i^2 (h, s))^{3/2} -( M_i^2 (h, s, T))^{3/2} \right\} \; . \label{d}
\end{eqnarray}
Thermal masses are given as $M_i^2 (h, s, T) = m_i^2(h, s) + \Pi_i (T^2)$, with $\Pi_i (T^2 )$ given in Eqs. (\ref{themalma}), (\ref{themalmb}) and (\ref{themalmc}).

We first study the zero temperature vacuum structures, which influence the electroweak phase transition at the finite temperature. The critical points are found by solving the minimization conditions
\begin{eqnarray}
\left.{\partial V_0 \over \partial h} \right|_{h_0}= \left.{\partial V_0 \over \partial s}\right |_{s_0} =0 \; ,
\end{eqnarray} 
which has at most nine solutions: $(0,~0)$, $(\pm v, ~\pm v_s)$, $(0, \pm \sqrt{\mu_s^2/\lambda_1})$ and $(\pm\sqrt{\mu^2 /\lambda}, ~0)$. There are four distinct critical points left after using the reflection symmetries to eliminate the redundant negative partners of these solutions.  We take $(v,~v_s)$ as the physical electroweak vacuum, where the scalar masses are given in the Eq. (\ref{eigenvalue}). The requirement of the vacuum stability may be summarized by the condition that this point is the global minimum. A naive calculation turns out that it always be true, if the solutions given in Eq. (\ref{vev})  are positive.  The requirement  can be written as the inequalities
\begin{eqnarray}
\lambda (\lambda_1) >0 \; ,\hspace{1cm}
4 \lambda \lambda_1 - \lambda_2^2 >0 \;, \hspace{1cm}  
2 \mu^2 \lambda_1 - \mu_s^2 \lambda_2 > 0 \; , \hspace{1cm}
2\mu_s^2 \lambda - \mu^2 \lambda_2 >0 \; .
\end{eqnarray}
which put constraints on the parameter space of the potential.  On the other hand, the vacuum stability and perturabativity~\cite{Chao:2012mx} of the SM Higgs at the high energy scale also constrain the parameter space.

We now derive conditions on the parameters such that the condition for a strongly first order EWPT obtains. Before doing so, we comment on the issue of gauge dependence of the EWPT. The root of the problem lies in the lack of gauge-invariant definition of the free energy.  Although the problem has not been  solved yet, there are some possible ways out.  A gauge independent condition for a strongly first order EWPT can be obtained in perturbative theory by  using a gauge invariant source term $j\Phi^\dagger \Phi$~\cite{Buchmuller:1994vy} in the generating functional. Another approach is working with a source term $j\Phi$,  that is not gauge invariant, and consistently implementing the Nielsen's identity~\cite{Nielsen:1975fs} and doing $\hbar $-expansion~\cite{Patel:2011th} with the effective potential, so as to erase the gauge parameter order by order.  In this paper we follow the approach given in Ref.\cite{Patel:2011th} to calculate the condition for the strongly first order EWPT. 

The requirement of an initially produced baryon asymmetry not to be washed out,  implies roughly a requirement on $\Delta E_{\rm sph }/ T_C$, that can be translated into the bound: $\phi({T_C})/ T_C \geq1.0$.  Although $\phi(T_C)$ is gauge dependent, one can obtain the sphaleron rate by evaluating the temperature-dependent effective action of the sphaleron where only the gauge independent ${\cal O} (T^2)$ terms are included. To this approximation,  the theory contains a gauge invariant  energy scale $\bar v(T)$, and the condition for a strongly first order EWPT turns out to be~\cite{Patel:2011th,Patel:2012pi}
\begin{eqnarray}
{ \bar v (T_C) \over T_C}\geq 1.0 \; ,
\end{eqnarray}
which is quoted as the criteria in our analysis of the EWPT.
\begin{figure}[t]
\centering
\includegraphics[width=0.6\textwidth]{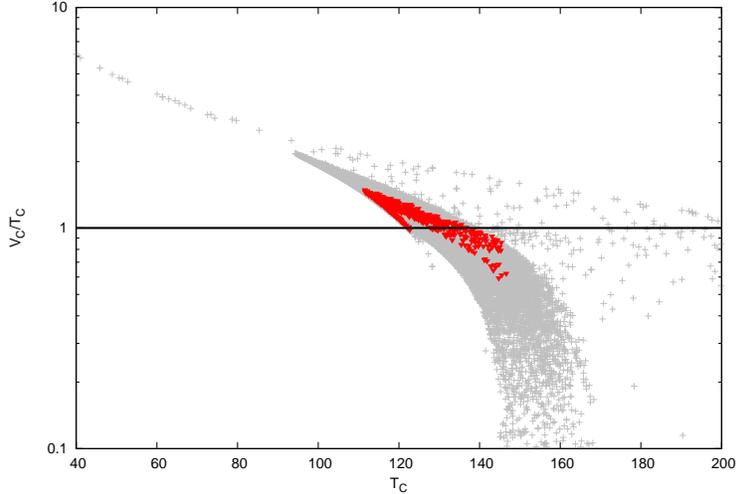}
\caption{ $\bar v(T_C)/T_C$ as the function of the critical temperature. Points in red color give a correct dark matter relic density and satisfy the LUX constraint in the meanwhile.  Points in gray are inconsistent with the constraint of dark matter observables. }           
\label{ptct}
\end{figure}

In the spirit of maintaining gauge independence, the effective potential,  in which only the gauge independent  ${\cal O} (T^2)$ terms are included, can be written as
\begin{eqnarray}
V(h,s,T)&=& {1 \over 2 }\left[  \Pi_h-  \mu_h^2   \right] h^2 +{1 \over 2 }\left[\Pi_s  -\mu_s^2  \right] s^2  + {1 \over 4 }  \lambda h^4 + {1 \over 4 } \lambda_1 s^4 + {1 \over 4 } h^2 s^2 \; , \label{tpotential}
\end{eqnarray}
where $\Pi_h$ and $\Pi_s$ are the thermal masses of the SM Higgs and scalar singlet respectively.  The sphaleron rate can be obtained from the effective action in Eq. (\ref{tpotential}) by performing path integral over the Higgs field.  The temperature dependent vacuum expectation value can be written as
\begin{eqnarray}
\bar v(T) = \sqrt{ v_0^2 +{2 \lambda_2 \Pi_s -4 \lambda_1 \Pi_h \over 4 \lambda \lambda_1 -\lambda_2^2 }  }
\end{eqnarray}
where $v_0 $ is the tree level VEV of the SM Higgs at the zero temperature. Notice that $\bar{v}(T)$ minimizes  $V(h,s,T)$ only for $ \Pi_h < \mu_h^2 $.  

We use the method given in Ref.~\cite{Patel:2011th} to calculate the critical temperature, by inserting the tree level minimas into the one-loop temperature-dependent effective potential, at which point the gauge dependence cancels. The critical temperature can be obtained by requiring  the following degeneracy condition
\begin{eqnarray}
V(h_1, s_1, T_C) = V( h_2, s_2, T_C) \; ,
\end{eqnarray}
 where $(h_1, ~s_1) =(0, \sqrt{\mu_s^2 / \lambda_1 })$ and $(h_2, s_2 ) =(246 , ~v_s)$. 


For the numerical analysis, we set $v_s$, $\lambda_1$ and $\lambda_2$ as  free parameters varying in the following ranges $v_s\in[ 100~{\rm GeV},~2~{\rm TeV}]$, $\lambda_1 \in (0,~2]$ and $ \lambda_2 \in (0,~1]$, the mass  and VEV of the SM Higgs are set to be $125~{\rm GeV}$  and $246~{\rm GeV}$ respectively, the dark matter mass is set to be near $m_s/2$, where $m_s$ is the mass of the scalar singlet at the zero temperature. All the other physical parameters can be obtained using these inputs.  In  Fig. \ref{ptct}, we plot  $\bar v(T_C) / T_C$ as the function of the critical temperature. Points in red color correspond to cases where the dark matter relic densities are consistent with the experimental observation within three standard deviations, and the LUX exclusion limits are satisfied.  One can see that the critical temperature lies in the range $[40,~200]~{\rm GeV}$.  For the points that satisfy all the constraints of dark matter observables the critical temperature roughly lies in the range$[110,~150]$.  
It is obvious that there are parameter space where  both strongly first order EWPT and the correct dark matter phenomenologies can be achieved. It should be mentioned that the barrier of the effective potential in this model comes from the thermal loop corrections.

\begin{figure}[t]
\centering
\includegraphics[width=0.45\textwidth]{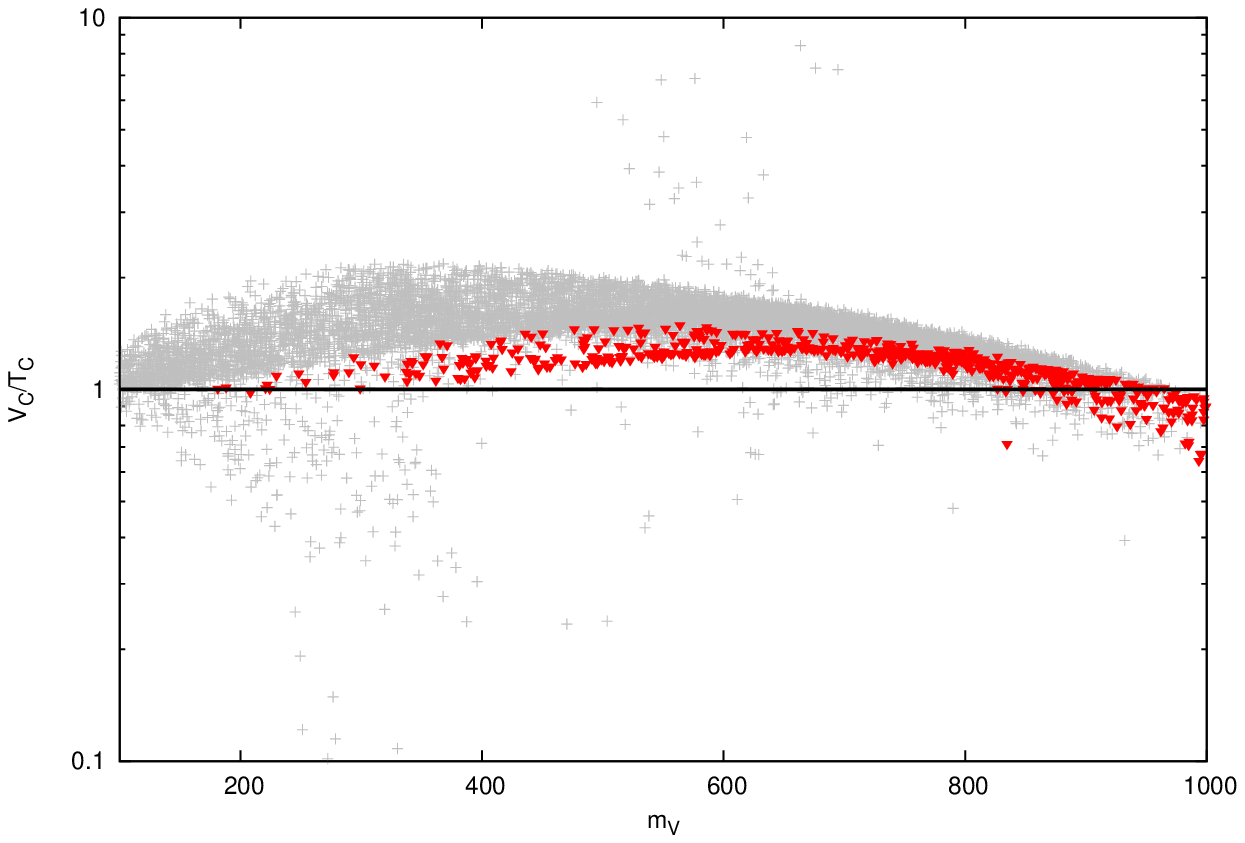}
\hspace{0.5cm}
\includegraphics[width=0.45\textwidth]{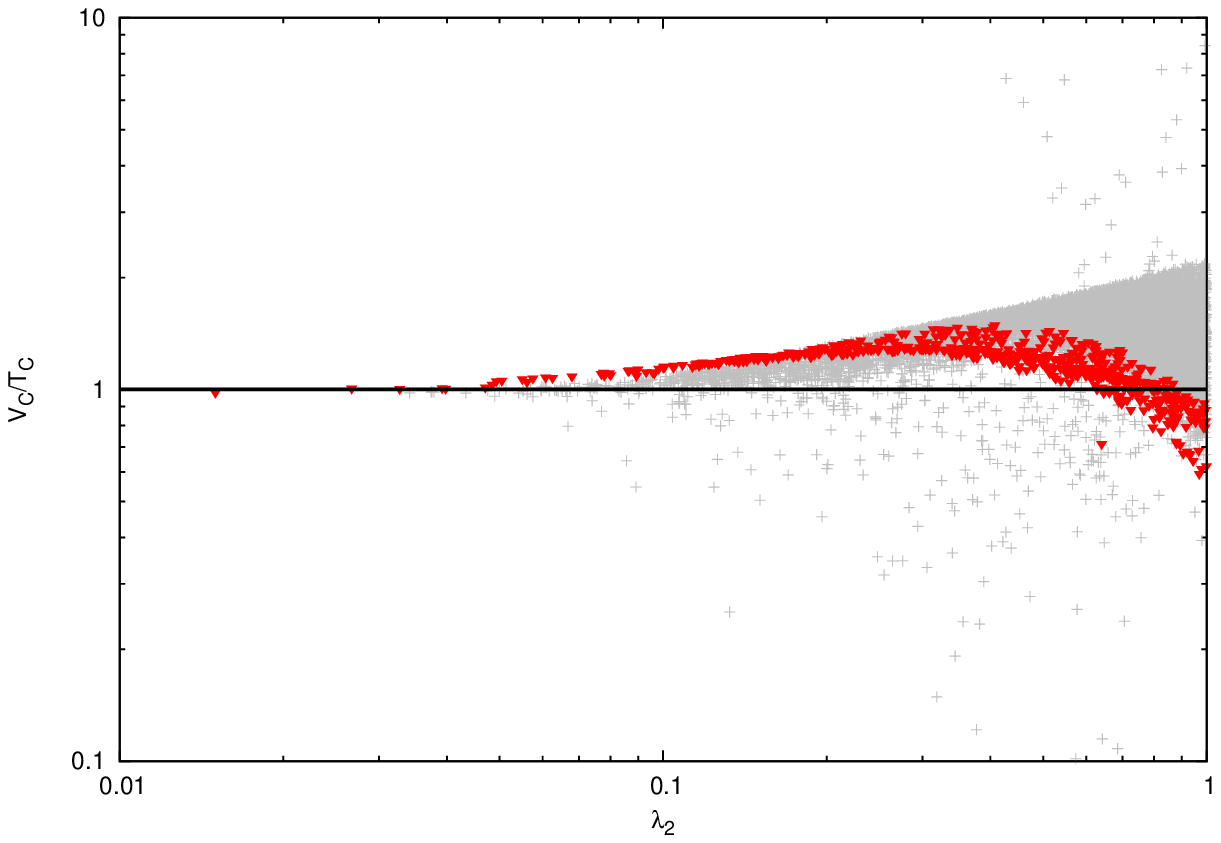}
\caption{ $\bar v(T_C)/T_C$ as the function of the dark matter mass(left panel) and as the function of the coupling $ \lambda_2$ (right panel). All the points in red color give a correct dark matter relic density and satisfy the LUX constraint in the meanwhile. The points in gray color are excluded by  the LUX.
 }\label{ptdm}
\end{figure}

In Fig. \ref{ptdm}, we plot $\bar v(T_C) /T_C$ as the the function of the dark matter mass (left panel) and the coupling $\lambda_2$ (right panel).  Points in red color (in reversed triangle symbol) satisfy all the dark matter constraints.  Points in gray color (in cross symbol) are excluded by the LUX. For a light dark matter the quartic coupling, $\lambda_2$, is constrained to be very small to give a correct dark matter relic density, which decreases the barrier of the effective potential induced by the same quartic coupling. As a result, one can not get a strongly first order EWPT in this case. For a much heavy dark matter, one  may have a large quartic coupling, $\lambda_2$, which can  be ${\cal O } (1)$, but the extra scalar will be too heavy and  will decouple when EWPT happens.  It should be mentioned that the gauge invariant EWPT method provided by ~\cite{Patel:2011th} somehow underestimates the critical temperature, which means the strength of the EWPT is overestimated. One needs to include higher order corrections to the effective potential to get a more accurate  result. Unfortunately, we are not aware of any computation of high order corrections to the effective potential in an arbitrary gauge.  Here we only give a rough estimation on the strength of the EWPT. From Fig.~\ref{ptdm}, $({\phi_C /T_C})_{\rm max} \approx 1.5$. We leave the study of the impact on the EWPT from the  next leading order  thermal corrections at ${\cal O} (\hbar^2)$ to a future work. Since the model we studied may not be the only new physics beyond the SM, we will not consider constraints of the oblique parameters on the model in this paper.

\section{Concluding remarks}

The discovery of the SM Higgs makes the EWPT  realistic. However the dynamic of the EWPT and its possible signatures are still unknown.  In this paper we have explored the parameter space of the Higgs portal vector dark matter model which can lead to strongly first order EWPT as required by the electroweak baryogenesis. We have studied the parameter space of the Higgs portal vector dark matter constrained by the LUX and the parameter space for a strongly first order EWPT. Our result shows that there are considerable parameter space where both successful dark matter phenomenologies and the first order EWPT can be achieved.  Our research are new in two aspects: (1) we perform a totally gauge invariant treatment of the EWPT;  (2) the model is closely related to the Higgs portal vector dark matter, which is distinctive compared with other Higgs portal dark matter scenarios, since the dark matter mass in our model is totally induced by the spontaneously broken of the $U(1)^\prime$ gauge symmetry.  The collider signatures of the model, which is interesting but beyond the scope of this study, will be shown in somewhere else.

\begin{acknowledgments}
The author thanks  Huaike Guo,  Michael Ramsey-Musolf  and Peter Winslow for useful discussions and comments on the manuscript.
This work was supported in part by DOE Grant DE-SC0011095.
\end{acknowledgments}


\begin{thebibliography}{99}

\bibitem{atlas} 
  G.~Aad {\it et al.}  [ATLAS Collaboration],
  Phys.\ Lett.\ B
  [arXiv:1207.7214 [hep-ex]].
  
  
\bibitem{Aad:2013wqa} 
  G.~Aad {\it et al.}  [ATLAS Collaboration],
  Phys.\ Lett.\ B {\bf 726}, 88 (2013)
  [arXiv:1307.1427 [hep-ex]].
  
  
\bibitem{cms} 
  S.~Chatrchyan {\it et al.}  [CMS Collaboration],
  Phys.\ Lett.\ B
  [arXiv:1207.7235 [hep-ex]].
  
\bibitem{Chatrchyan:2013lba} 
  S.~Chatrchyan {\it et al.}  [CMS Collaboration],
  JHEP {\bf 1306}, 081 (2013)
  [arXiv:1303.4571 [hep-ex]].
  
  \bibitem{higgs}

F. Englert and R. Brout, Phys. Rev. Lett. {\bf 13}, 321 (1964); P. W. Higgs, Phys. Rev. Lett. {\bf 12}, 132 (1964); Phys. Rev. Lett. {\bf 13}, 508(1964); Phys. Rev.  {\bf 145}, 1156 (1966); G. S. Guralnik, Phys. Rev. Lett. {\bf 13}, 585(1964); T. W. B. Kibble, Phys. Rev. {\bf 155}, 1554 (1967).
  


\bibitem{Kirzhnits:1972ut} 
  D.~A.~Kirzhnits and A.~D.~Linde,
  Phys.\ Lett.\ B {\bf 42}, 471 (1972).


\bibitem{Rubakov:1996vz} 
  V.~A.~Rubakov and M.~E.~Shaposhnikov,
  Usp.\ Fiz.\ Nauk {\bf 166}, 493 (1996)
  [Phys.\ Usp.\  {\bf 39}, 461 (1996)]
  [hep-ph/9603208].


\bibitem{Morrissey:2012db} 
  D.~E.~Morrissey and M.~J.~Ramsey-Musolf,
  New J.\ Phys.\  {\bf 14}, 125003 (2012)
  [arXiv:1206.2942 [hep-ph]].
  
  
\bibitem{Quiros:1999jp} 
  M.~Quiros,
  hep-ph/9901312.
  
  
  
\bibitem{Ade:2013zuv} 
  P.~A.~R.~Ade {\it et al.}  [Planck Collaboration],
  Astron.\ Astrophys.\  (2014)
  [arXiv:1303.5076 [astro-ph.CO]].
  

  
\bibitem{dmn1}
G. Bertone, D. Hooper and J. Silk, Phys. Rept. {\bf 405}, 279 (2005).

\bibitem{dmn2}

G. Jungman, M. Kamionkowski and K. Griest, Phys. Rept. {\bf 267}, 195 (1996)

\bibitem{dmn3}

G. Bertone and (ed.), {\it Particle dark matter: Observations, models and searches.  } Cambridge 
University Press, 2010.


  
\bibitem{Patt:2006fw} 
  B.~Patt and F.~Wilczek,
  hep-ph/0605188.
  
\bibitem{Kim:2006af} 
  Y.~G.~Kim and K.~Y.~Lee,
  Phys.\ Rev.\ D {\bf 75}, 115012 (2007)
  [hep-ph/0611069].
  
  
\bibitem{MarchRussell:2008yu} 
  J.~March-Russell, S.~M.~West, D.~Cumberbatch and D.~Hooper,
  JHEP {\bf 0807}, 058 (2008)
  [arXiv:0801.3440 [hep-ph]].

\bibitem{Kim:2008pp} 
  Y.~G.~Kim, K.~Y.~Lee and S.~Shin,
  JHEP {\bf 0805}, 100 (2008)
  [arXiv:0803.2932 [hep-ph]].
  
\bibitem{Ahlers:2008qc} 
  M.~Ahlers, J.~Jaeckel, J.~Redondo and A.~Ringwald,
  Phys.\ Rev.\ D {\bf 78}, 075005 (2008)
  [arXiv:0807.4143 [hep-ph]].
  
  
\bibitem{Feng:2008mu} 
  J.~L.~Feng, H.~Tu and H.~-B.~Yu,
  JCAP {\bf 0810}, 043 (2008)
  [arXiv:0808.2318 [hep-ph]].
  
\bibitem{Andreas:2008xy} 
  S.~Andreas, T.~Hambye and M.~H.~G.~Tytgat,
  JCAP {\bf 0810}, 034 (2008)
  [arXiv:0808.0255 [hep-ph]].
  
\bibitem{Barger:2008jx} 
  V.~Barger, P.~Langacker, M.~McCaskey, M.~Ramsey-Musolf and G.~Shaughnessy,
  Phys.\ Rev.\ D {\bf 79}, 015018 (2009)
  [arXiv:0811.0393 [hep-ph]].
  
\bibitem{Kadastik:2009ca} 
  M.~Kadastik, K.~Kannike, A.~Racioppi and M.~Raidal,
  Phys.\ Rev.\ Lett.\  {\bf 104}, 201301 (2010)
  [arXiv:0912.2729 [hep-ph]].
  
  
\bibitem{Piazza:2010ye} 
  F.~Piazza and M.~Pospelov,
  Phys.\ Rev.\ D {\bf 82}, 043533 (2010)
  [arXiv:1003.2313 [hep-ph]].
  
\bibitem{Arina:2010an} 
  C.~Arina, F.~-X.~Josse-Michaux and N.~Sahu,
  Phys.\ Rev.\ D {\bf 82}, 015005 (2010)
  [arXiv:1004.3953 [hep-ph]].
  

\bibitem{Englert:2011yb} 
  C.~Englert, T.~Plehn, D.~Zerwas and P.~M.~Zerwas,
  Phys.\ Lett.\ B {\bf 703}, 298 (2011)
  [arXiv:1106.3097 [hep-ph]].
  
\bibitem{Low:2011kp} 
  I.~Low, P.~Schwaller, G.~Shaughnessy and C.~E.~M.~Wagner,
  Phys.\ Rev.\ D {\bf 85}, 015009 (2012)
  [arXiv:1110.4405 [hep-ph]].
  

  
  
\bibitem{Kamenik:2012hn} 
  J.~F.~Kamenik and C.~Smith,
  Phys.\ Rev.\ D {\bf 85}, 093017 (2012)
  [arXiv:1201.4814 [hep-ph]].
  
\bibitem{Gonderinger:2012rd} 
  M.~Gonderinger, H.~Lim and M.~J.~Ramsey-Musolf,
  Phys.\ Rev.\ D {\bf 86}, 043511 (2012)
  [arXiv:1202.1316 [hep-ph]].
  
\bibitem{Lebedev:2012zw} 
  O.~Lebedev,
  Eur.\ Phys.\ J.\ C {\bf 72}, 2058 (2012)
  [arXiv:1203.0156 [hep-ph]].
  
\bibitem{LopezHonorez:2012kv} 
  L.~Lopez-Honorez, T.~Schwetz and J.~Zupan,
  Phys.\ Lett.\ B {\bf 716}, 179 (2012)
  [arXiv:1203.2064 [hep-ph]].
  
  
\bibitem{Okada:2012cc} 
  H.~Okada and T.~Toma,
  Phys.\ Lett.\ B {\bf 713}, 264 (2012)
  [arXiv:1203.3116 [hep-ph]].
  
\bibitem{Djouadi:2012zc} 
  A.~Djouadi, A.~Falkowski, Y.~Mambrini and J.~Quevillon,
  Eur.\ Phys.\ J.\ C {\bf 73}, 2455 (2013)
  [arXiv:1205.3169 [hep-ph]].
  
\bibitem{Bai:2012nv} 
  Y.~Bai, V.~Barger, L.~L.~Everett and G.~Shaughnessy,
  Phys.\ Rev.\ D {\bf 88}, no. 1, 015008 (2013)
  [arXiv:1212.5604 [hep-ph]].
  
  
\bibitem{Chao:2012pt} 
  W.~Chao and M.~J.~Ramsey-Musolf,
  Phys.\ Rev.\ D {\bf 89}, no. 3, 033007 (2014)
  [arXiv:1212.5709 [hep-ph]].
  
\bibitem{Englert:2013gz} 
  C.~Englert, J.~Jaeckel, V.~V.~Khoze and M.~Spannowsky,
  JHEP {\bf 1304}, 060 (2013)
  [arXiv:1301.4224 [hep-ph]].
  
\bibitem{Bian:2013wna} 
  L.~Bian, R.~Ding and B.~Zhu,
  Phys.\ Lett.\ B {\bf 728}, 105 (2014)
  [arXiv:1308.3851 [hep-ph]].
  
\bibitem{Chang:2013lfa} 
  C.~-F.~Chang, E.~Ma and T.~-C.~Yuan,
  JHEP {\bf 1403}, 054 (2014)
  [arXiv:1308.6071 [hep-ph], arXiv:1308.6071].
  
\bibitem{Khoze:2013uia} 
  V.~V.~Khoze,
  JHEP {\bf 1311}, 215 (2013)
  [arXiv:1308.6338 [hep-ph]].
  
\bibitem{Okada:2013bna} 
  N.~Okada and O.~Seto,
  Phys.\ Rev.\ D {\bf 89}, 043525 (2014)
  [arXiv:1310.5991 [hep-ph]].
  

\bibitem{Fedderke:2014wda} 
  M.~A.~Fedderke, J.~-Y.~Chen, E.~W.~Kolb and L.~-T.~Wang,
  arXiv:1404.2283 [hep-ph].
  
\bibitem{Djouadi:2011aa} 
  A.~Djouadi, O.~Lebedev, Y.~Mambrini and J.~Quevillon,
  Phys.\ Lett.\ B {\bf 709}, 65 (2012)
  [arXiv:1112.3299 [hep-ph]].


\bibitem{Kanemura:2010sh} 
  S.~Kanemura, S.~Matsumoto, T.~Nabeshima and N.~Okada,
  Phys.\ Rev.\ D {\bf 82}, 055026 (2010)
  [arXiv:1005.5651 [hep-ph]].
  
  


\bibitem{Baek:2012se} 
  S.~Baek, P.~Ko, W.~I.~Park and E.~Senaha,
  JHEP {\bf 1305}, 036 (2013)
  [arXiv:1212.2131 [hep-ph]].

\bibitem{Lebedev:2011iq} 
  O.~Lebedev, H.~M.~Lee and Y.~Mambrini,
  Phys.\ Lett.\ B {\bf 707}, 570 (2012)
  [arXiv:1111.4482 [hep-ph]].










\bibitem{Pietroni:1992in} 
  M.~Pietroni,
  Nucl.\ Phys.\ B {\bf 402}, 27 (1993)
  [hep-ph/9207227].

\bibitem{Espinosa:1993bs} 
  J.~R.~Espinosa and M.~Quiros,
  Phys.\ Lett.\ B {\bf 305}, 98 (1993)
  [hep-ph/9301285].
  
\bibitem{John:1998ip} 
  P.~John,
  Phys.\ Lett.\ B {\bf 452}, 221 (1999)
  [hep-ph/9810499].

\bibitem{Huber:2000mg} 
  S.~J.~Huber and M.~G.~Schmidt,
  Nucl.\ Phys.\ B {\bf 606}, 183 (2001)
  [hep-ph/0003122].
  
\bibitem{Ham:2004cf} 
  S.~W.~Ham, Y.~S.~Jeong and S.~K.~Oh,
  J.\ Phys.\ G {\bf 31}, 857 (2005)
  [hep-ph/0411352].
  
\bibitem{Profumo:2007wc} 
  S.~Profumo, M.~J.~Ramsey-Musolf and G.~Shaughnessy,
  JHEP {\bf 0708}, 010 (2007)
  [arXiv:0705.2425 [hep-ph]].
  
\bibitem{Noble:2007kk} 
  A.~Noble and M.~Perelstein,
  Phys.\ Rev.\ D {\bf 78}, 063518 (2008)
  [arXiv:0711.3018 [hep-ph]].
  
  
\bibitem{Espinosa:2011ax} 
  J.~R.~Espinosa, T.~Konstandin and F.~Riva,
  Nucl.\ Phys.\ B {\bf 854}, 592 (2012)
  [arXiv:1107.5441 [hep-ph]].
  
  
  
  
  
\bibitem{Cline:2012hg} 
  J.~M.~Cline and K.~Kainulainen,
  JCAP {\bf 1301}, 012 (2013)
  [arXiv:1210.4196 [hep-ph]].
  
\bibitem{Cline:2013gha} 
  J.~M.~Cline, K.~Kainulainen, P.~Scott and C.~Weniger,
  Phys.\ Rev.\ D {\bf 88}, 055025 (2013)
  [arXiv:1306.4710 [hep-ph]].
  
\bibitem{Katz:2014bha} 
  A.~Katz and M.~Perelstein,
  JHEP {\bf 1407}, 108 (2014)
  [arXiv:1401.1827 [hep-ph]].
  
  
\bibitem{Alanne:2014bra} 
  T.~Alanne, K.~Tuominen and V.~Vaskonen,
  arXiv:1407.0688 [hep-ph].
  
  
  
  
\bibitem{Profumo:2014opa} 
  S.~Profumo, M.~J.~Ramsey-Musolf, C.~L.~Wainwright and P.~Winslow,
  arXiv:1407.5342 [hep-ph].


\bibitem{Curtin:2014jma} 
  D.~Curtin, P.~Meade and C.~T.~Yu,
  arXiv:1409.0005 [hep-ph].





















\bibitem{Akerib:2013tjd} 
  D.~S.~Akerib {\it et al.}  [LUX Collaboration],
  Phys.\ Rev.\ Lett.\  {\bf 112}, 091303 (2014)
  [arXiv:1310.8214 [astro-ph.CO]].


  
  


  
\bibitem{Buchmuller:1994vy} 
  W.~Buchmuller, Z.~Fodor and A.~Hebecker,
  Phys.\ Lett.\ B {\bf 331}, 131 (1994)
  [hep-ph/9403391].
  
  
  
  
\bibitem{Nielsen:1975fs} 
  N.~K.~Nielsen,
  Nucl.\ Phys.\ B {\bf 101}, 173 (1975).
  
  
\bibitem{Patel:2011th} 
  H.~H.~Patel and M.~J.~Ramsey-Musolf,
  JHEP {\bf 1107}, 029 (2011)
  [arXiv:1101.4665 [hep-ph]].
  
  
\bibitem{Patel:2012pi} 
  H.~H.~Patel and M.~J.~Ramsey-Musolf,
  Phys.\ Rev.\ D {\bf 88}, 035013 (2013)
  [arXiv:1212.5652 [hep-ph]].
  
  

  
\bibitem{Chao:2012mx} 
  W.~Chao, M.~Gonderinger and M.~J.~Ramsey-Musolf,
  Phys.\ Rev.\ D {\bf 86}, 113017 (2012)
  [arXiv:1210.0491 [hep-ph]].
  
  
\end{thebibliography}
\end{document}